\documentclass[final,12pt]{article}

\usepackage[letterpaper,margin=1in]{geometry}
\usepackage{amsfonts}
\usepackage{color}
\usepackage{amssymb}
\usepackage{amsmath}
\usepackage{amsthm}
\usepackage{bbm}
\usepackage{enumerate}
\usepackage{fancyhdr}
\usepackage{setspace}
\usepackage{abstract}
\usepackage{natbib}
\usepackage{rotating}
\usepackage{multicol}
\usepackage{overpic}
\usepackage[colorlinks=true, citecolor=black, linkcolor=darkblue, pagebackref=false, breaklinks=true]{hyperref}
\usepackage{booktabs}
\usepackage{threeparttable}
\usepackage{float}
\usepackage{titlesec}
\usepackage{caption}
\usepackage{leftidx}
\usepackage{mathabx}
\usepackage[page,title]{appendix}
\usepackage{chngcntr}
\usepackage{tikz}
\usepackage{subcaption}

\theoremstyle{definition}

\definecolor{darkblue}{rgb}{0,0,0.55}
\definecolor{lightgray}{rgb}{0.65,0.65,0.65}

\newcommand{\nin}{\noindent}
\newcommand\Bheadfont{\fontsize{14pt}{\baselineskip}\selectfont}

\numberwithin{equation}{section}

\titleformat{\section}[hang]{\normalfont\sc\color{darkblue}\Bheadfont} {\thesection\hskip0.618em}{0em}{}
\titlespacing*{\section}{0pt}{15pt plus 2pt minus 2pt}{9pt plus 2pt minus 2pt}
\titleformat{\subsection}[runin]{\normalfont\sc\color{darkblue}} {\thesubsection\hskip0.618em}{0em}{}
\titlespacing*{\subsection}{0pt}{13pt plus 2pt minus 2pt}{13pt plus 2pt minus 2pt}
\begin{document}
\renewcommand\footnotemark{}

\title{\LARGE Learning the Macroeconomic Language}

\author{Siddhartha Chib, Fei Tan, and Zhixun Zhang$^{\Asterisk}$\thanks{\hspace*{-0.2cm}$^{\Asterisk}$Chib: Olin Business School, Washington University in St. Louis; Tan: Chaifetz School of Business, Saint Louis University and Center for Economic Behavior and Decision-Making, Zhejiang University of Finance and Economics; Zhang: Department of Economics, University of Chicago. Send correspondence to email: \href{mailto:tanf@slu.edu}{\color{blue}tanf@slu.edu} (F. Tan).} \date{\today}}

\maketitle
\vspace{-5mm}

\begin{abstract}
    \normalsize   
    
    \noindent We show how state-of-the-art large language models (LLMs) can be trained effectively on limited historical data for macroeconomic forecasting. We estimate a dynamic stochastic general equilibrium (DSGE) model with stochastic volatility and Student-$t$ shocks on an initial segment of the data to obtain a posterior distribution over structural parameters. We sample from this posterior to generate millions of theory-consistent synthetic panels that, when mixed with macroeconomic data, form the training corpus for a deep learning transformer. Rather than using the DSGE model as a direct forecasting device, we use it as a structural simulator that regularizes transformer training in a small-sample environment. Our results show that this hybrid forecaster, which combines the theoretical coherence of DSGE models with the representational power of modern LLMs, learns key features of the macroeconomic language. Relative to a conventional vector autoregression benchmark, the transformer achieves comparable or stronger predictive performance in both token accuracy and predictive likelihood. These gains are strongest under a theory-heavy training mix and remain broadly robust to finer tokenization and deeper network architecture.\\

    \nin {\bf\it Keywords}: Bayesian methods; Deep learning; DSGE models; Economic forecasting; Large language models; Transformers.\\

    \nin {\bf\it JEL Classification}: C11, C45, E37
\end{abstract}

\thispagestyle{empty}
\newpage
\setcounter{page}{1}

\section{Introduction}\label{sec:intro}

\doublespacing

Macroeconomic forecasting has long confronted a fundamental challenge, that of learning reliable dynamic relationships from limited historical data. Quarterly macroeconomic time series rarely exceed a few hundred observations, a sample size that is far too small to support the direct application of modern machine learning methods. In contrast, recent advances in sequence modeling, most notably the transformer architecture of \cite{attention}, have been driven by training on billions of examples. The question we address is whether these powerful sequence learners can be adapted to the inherently small-sample environment of macroeconomics.

A recent illustration of the flexibility of transformer architectures is provided by the Starcloud-1 space station, where an NVIDIA H100 was used to train the nanoGPT model of \cite{nanoGPT} on Shakespeare's complete works, the first large language model (LLM) trained in orbit. This demonstration shows that transformers, once confined to massive data centers, can be deployed under extreme computational constraints through careful architectural design and training strategies. Adapting these methods to macroeconomic forecasting, however, presents a different challenge, that is the scarcity of data rather than computational capacity.

In this paper, we show how transformer models can be trained effectively for macroeconomic forecasting by combining two complementary sources of information, limited historical observations and abundant theory-consistent synthetic data that are generated from estimated dynamic stochastic general equilibrium (DSGE) models. The key insight is to use DSGE models not as direct forecasting devices, but as structured data generators that encode economic theory. In this respect, our approach is loosely related to the literature where DSGE models are used to inform priors in reduced-form time series models [see \cite{bc-var1}, \cite{bc-var2}, \cite{dsge-var}, and \cite{sw-var}]. Beyond this broad conceptual connection, our method is fundamentally different in its objectives, implementation, and intended application to transformer-based sequence learning.

We proceed in two steps. First, we estimate a \cite{SW07} style DSGE model augmented with stochastic volatility and Student-$t$ shocks using Bayesian methods on an initial segment of historical data. This estimation yields a posterior distribution over structural parameters that captures uncertainty about the economy's underlying dynamics. We then draw from this posterior to generate millions of synthetic macroeconomic trajectories. Each trajectory represents a plausible counterfactual realization of the economy and incorporates empirically relevant features such as time-varying volatility and fat-tailed shocks. Second, we train a transformer model using batches that combine synthetic observations with real historical data. This mixed training strategy allows the model to absorb economic structure from DSGE-based simulations while remaining anchored to observed macroeconomic outcomes. The synthetic trajectories expose the transformer to a wide range of economic scenarios, including regimes and crisis episodes that are sparsely represented, or entirely absent, in historical samples. In this way, the procedure directly addresses the small-sample problem that limits the applicability of modern machine learning methods in macroeconomics.

Our architectural design incorporates two adaptations that are essential for macroeconomic forecasting. The first concerns the multivariate nature of macroeconomic data. Unlike standard language models, which process a single token at a time, macroeconomic forecasting requires handling several aggregate variables that are observed jointly each period. A naive joint tokenization of all variables would lead to a combinatorial explosion in the vocabulary, rendering the approach infeasible even for modest discretizations. To address this issue, we adopt a \emph{separate-then-concatenate} embedding strategy in which each variable is tokenized and embedded separately, and the resulting embeddings are concatenated before entering the attention layers of the model. This construction reduces the number of parameters by several orders of magnitude while still allowing the attention mechanism to learn cross-variable dependencies in a flexible manner.

The second adaptation concerns the model's output. Drawing inspiration from modular systems in robotics and autonomous driving, rather than estimating a single, monolithic transformer with a high-dimensional output layer, we employ a modular design consisting of seven specialized transformers, one for each target variable. Each transformer conditions on the same multivariate history but produces forecasts for a single macroeconomic series. This modularization avoids the curse of dimensionality in the output space and yields models that are both stable and computationally efficient. In practice, our baseline specification contains on the order of 50,000 parameters and can be trained in minutes on standard hardware, while retaining sufficient capacity to capture nonlinear dynamics and cross-variable interactions.

We also depart fundamentally from conventional transfer learning paradigms in which models are pretrained on a large dataset and subsequently fine tuned on a smaller one. In the present setting, such an approach would be ineffective. With only a few hundred quarterly observations relative to the volume of synthetic data, fine-tuning would leave the pretrained representations essentially unchanged, causing the information in the real data to be largely ignored. Instead, we implement \textit{mixed-batch training} wherein each training iteration samples from both real and synthetic data in a controlled mixture (10\% real vs. 90\% synthetic in our baseline). This ensures that empirical data influence learning from the outset, while synthetic data stabilize estimation and mitigate overfitting.

The mixing proportion has an interpretation analogous to prior strength in Bayesian inference. Synthetic data encode theoretical restrictions implied by the DSGE model, while real data provide direct empirical evidence. The training process balances these two sources of information in a transparent and controllable manner. More broadly, this approach illustrates how simulation-based models and observational data can be combined within modern deep-learning frameworks, with potential applications beyond macroeconomics to any domain where structural simulators exist but real data are scarce.

Our approach connects to a growing body of work in which simulation-based training is used to overcome data scarcity. In autonomous driving, agents are trained extensively in physics-based simulators such as CARLA \citep{car} before deployment in real environments. In robotics, synthetic manipulation data generated by simulation engines are routinely used to train control policies, with domain randomization techniques \citep{robot} enabling successful transfer from simulated to real-world environments. In each case, the simulator encodes domain-specific structure that allows learning to proceed efficiently despite limited real-world observations. In macroeconomics, DSGE models play an analogous role. They provide theory-consistent simulators that generate economically coherent trajectories, making it possible to train modern transformer-based models even when historical data are scarce.

We apply the method to forecast seven key U.S. macroeconomic variables over an out-of-sample period spanning nearly eight years. The results demonstrate strong predictive performance across all variables. When prediction errors occur, they are typically small; the model selects adjacent tokens rather than making large forecasting mistakes, which suggests that the learned representations capture economically meaningful relationships. The model performs particularly well on persistent level variables (hours worked, inflation, interest rate), consistent with their stronger autocorrelation and mean reversion. Under the baseline specification with a strong DSGE prior, the transformer matches or exceeds a rolling vector autoregression (VAR) benchmark in token accuracy for all variables and improves predictive log likelihood for most series. Additional robustness exercises show that weakening the DSGE prior degrades overall performance, whereas both finer tokenization and deeper network architecture leave the main distributional forecasting advantage largely intact.

The remainder of the paper is organized as follows. Section~\ref{sec:dsge} describes the DSGE model, data partition, and synthetic data generation. Section~\ref{sec:transformer} details the transformer architecture, embedding strategy, and mixed-batch training procedure. Section~\ref{sec:posterior} presents the empirical results, including forecast evaluation and robustness checks. Section~\ref{sec:conclusion} concludes the paper. The appendix contains a complete mathematical description of the transformer model used in this paper (Appendix \ref{app:transformer}) along with the Python implementation of our forecasting framework (Appendix \ref{app:code}).

\section{DSGE Prior}\label{sec:dsge}

The core idea of our approach rests on using a structural DSGE model as a data augmentation mechanism. This section describes how we estimate the DSGE parameters, partition the observed data, and generate synthetic training trajectories that encode economic theory while respecting parameter uncertainty.

\subsection{Model Setup}

We adopt the medium-scale DSGE model of \cite{SW07}, which provides a rich structural environment featuring (i) monopolistic competition in goods and labor markets with Calvo pricing and wage rigidities; (ii) investment adjustment costs and variable capital utilization; (iii) habit formation in consumption; (iv) seven structural shocks (technology, risk premium, investment-specific technology, government spending, price markup, wage markup, and monetary policy); and (v) seven observables (output growth, consumption growth, investment growth, wage growth, hours worked, inflation, and nominal interest rate).

Specifically, the Smets-Wouters model consists of 46 log-linearized equilibrium equations and can be expressed generically in the form
\begin{gather}
  \underset{(46\times 46)}{\Gamma_0(\theta^S)} x_t = \underset{(46\times 46)}{\Gamma_1(\theta^S)}x_{t-1} + \underset{(46\times 7)}{\Psi}\epsilon_t + \underset{(46\times 12)}{\Pi}\eta_t,\label{model}
\end{gather}
where $\theta^S$ is a $29\times 1$ vector of structural parameters, $x_t$ is a $46\times 1$ vector of model variables, $\epsilon_t$ is a $7\times 1$ vector of shock innovations, $\eta_t$ is a $12\times 1$ vector of forecast errors, and $(\Gamma_0,\Gamma_1,\Psi,\Pi)$ are coefficient matrices with dimensions indicated below them.

We make two model extensions to capture tail events and time-varying uncertainty observed in macroeconomic data, particularly during financial crises and economic recessions. First, the shock innovations follow a multivariate Student-$t$ distribution, i.e., $\epsilon_t\sim t_{\nu}(0,\Sigma_t)$. Here $\nu$ denotes the degrees of freedom and $\Sigma_t$ is a diagonal matrix with the time-varying volatility $\sigma_{i,t}^2$ for each individual innovation $\epsilon_t^i$ on its main diagonal, where $i$ refers to the shock index. For estimation convenience, it is useful to represent each element of $\epsilon_t$ as a mixture of normals by introducing a Gamma distributed random variable $\lambda_t$
\begin{gather}
  \epsilon_{t}^i=\lambda_t^{-1/2}e^{h_{t}^i/2}\varepsilon_{t}^i,\qquad\lambda_t\sim\mathbb{G}\left(\frac{\nu}{2},\frac{\nu}{2}\right),\qquad\varepsilon_{t}^i\sim\mathbb{N}(0,1).\label{gam_norm}
\end{gather}
Second, the logarithm of each volatility $h_{t}^i=\ln\sigma_{i,t}^2$ collected in a $7\times 1$ vector $h_t$ evolves as a stationary $(|\phi_i|<1)$ process
\begin{gather}
  h_{t}^i=(1-\phi_i)\mu_i+\phi_ih_{t-1}^i+\eta_{t}^i,\qquad\eta_{t}^i\sim\mathbb{N}(0,\omega_i^2),\label{sv}
\end{gather}
where we collect all the volatility parameters $(\mu_i,\phi_i,\omega_i^2)$ in a $21\times 1$ vector $\theta^V$. Taken together, these features are essential for generating realistic synthetic training data that expose the transformer to extreme economic scenarios rarely present in historical samples.

To conserve space, we refer the interested reader to \cite{SW07} for a detailed model description, data construction, and priors for the structural parameters. For the stochastic-volatility parameters, the unconditional means $\mu_i$ have diffuse Normal priors $\mathbb{N}(0,10)$ centered at zero on the log-volatility scale, the persistence parameters $\phi_i$ have tight Beta priors $\mathbb{B}(0.9,0.05)$ centered at $0.90$ to reflect the high persistence typically observed in macroeconomic volatility, and the innovation variances $\omega_i^2$ have Inverse-Gamma type-II priors $\mathbb{IG}\text{-}2(2,0.05)$ with small scale parameters that favor moderate volatility of volatility.

\subsection{Real Data}

Equations \eqref{model}--\eqref{sv} form a high-dimensional DSGE model that is completed with a measurement equation connecting the state variables $x_t$ to a $K\times 1$ vector of observable measurements $y_t$. Our quarterly U.S. macroeconomic dataset spans 1947:Q3 to 2025:Q2, comprising 312 observations of $K=7$ key variables as constructed in \cite{SW07}. We partition the data into three non-overlapping segments.

\begin{enumerate}
    \item \textbf{DSGE estimation sample} (50 observations, 1947:Q3--1959:Q4): Used exclusively for estimating the DSGE posterior distribution. This segment provides sufficient variation to identify structural parameters while remaining completely separate from transformer training and testing. Specifically, given this sample data $y_{1947:1959}$, the goal is to learn about (i) 50 model parameters $\theta = (\theta^S,\theta^V)$, (ii) 50 non-Gaussian latent variables $\lambda_{1947:1959}$ needed for modeling Student-$t$ shocks, and (iii) 350 nonlinear latent variables $h_{1947:1959}$ representing log volatilities. Let $p(\theta)$ denote the prior distribution. We follow the Bayesian estimation procedure detailed in \cite{dsge-svt} and obtain 10,000 posterior draws of the augmented posterior distribution
    \begin{gather*}
        p(\theta,\lambda_{1947:1959},h_{1947:1959}\mid y_{1947:1959})\ \propto\ f(y_{1947:1959},\lambda_{1947:1959},h_{1947:1959}\mid \theta)\cdot p(\theta)\cdot\mathbbm{1}\{\theta\in\Theta_D\},
    \end{gather*}
    where $f(y_{1947:1959},\lambda_{1947:1959},h_{1947:1959}\mid \theta)$ is the likelihood function, and $\mathbbm{1}\{\theta\in\Theta_D\}$ is an indicator function that equals one if $\theta$ is in the determinacy region $\Theta_D$ and zero otherwise.

    \item \textbf{Transformer training sample} (231 observations, 1960:Q1--2017:Q3): The real macroeconomic data that, when mixed with synthetic DSGE simulations, form the training corpus for the transformer. All data standardizations (means, standard deviations) are computed exclusively from this segment to prevent look-ahead bias.

    \item \textbf{Out-of-sample test sample} (31 observations, 2017:Q4--2025:Q2): The final 10\% of the sample spanning nearly eight years, completely held out for forecast evaluation. This segment includes major economic events (COVID-19 pandemic, inflation surge of 2021-2023) that stress-test the model's generalization capability.
\end{enumerate}

This partition strategy ensures that (i) DSGE estimation and transformer training use non-overlapping historical periods, providing independent information sources; and (ii) the transformer never sees test-period data during training.

\subsection{Synthetic Data}

Transformer training requires millions of observations, far exceeding the 231 quarters available in our training sample. We address this fundamental constraint by generating synthetic trajectories from the DSGE posterior predictive distribution. Crucially, rather than simulating at a single parameter value $\hat{\theta}$ (such as the posterior mean or mode), we implement proper Bayesian predictive sampling. For each synthetic trajectory $m = 1, \ldots, M$, we (i) draw structural parameters from the DSGE posterior, $\theta^{(m)} \sim p(\theta\mid y_{1947:1959})$; and (ii) simulate the DSGE model forward for $S$ periods using $\theta^{(m)}$, generating trajectory $\{y_t^{(m)}\}_{t=1}^S$. We generate $M = 10{,}000$ synthetic trajectories, each of length $S = 1{,}000$ quarters, yielding 10 million synthetic observations. This massive synthetic corpus dwarfs the 231 real training observations, enabling transformer training that would otherwise be infeasible.

The posterior predictive sampling offers several advantages. First, it embeds parameter uncertainty directly into the training data; the transformer experiences diverse economic dynamics corresponding to different plausible parameterizations rather than committing to a single calibration. Second, it generates regime diversity. By sampling across the posterior, we create trajectories exhibiting varying degrees of nominal rigidity, shock persistence, and volatility, exposing the transformer to economic environments spanning the uncertainty reflected in the DSGE posterior. Third, the stochastic volatility and Student-$t$ features ensure synthetic data include tail events and crises, preparing the model for forecasting during turbulent periods.

\subsection{Hierarchical Bayesian Interpretation}

Our procedure admits a clean hierarchical Bayesian interpretation. The DSGE estimation yields a posterior distribution that encodes structural beliefs about the economy. The synthetic trajectories drawn from this posterior constitute the prior predictive distribution for future macroeconomic outcomes. The transformer serves as a likelihood function that performs Bayesian updating. It starts with the DSGE prior and refines predictions using the empirical signal from real observations. The mixing ratio between synthetic and real data acts as a hyperparameter controlling the prior-to-likelihood weight, analogous to the effective sample size in Bayesian inference.

This framework contrasts with traditional DSGE forecasting, where predictions are generated via the Kalman filter. Instead, we use the DSGE model to generate a training corpus, then let the transformer learn a flexible forecasting function that need not respect the DSGE's linearized structure. The transformer can discover nonlinear patterns, time-varying relationships, and cross-variable dependencies that the DSGE's equilibrium conditions might misspecify, while still benefiting from the economic structure encoded in synthetic data.

Computationally, DSGE simulation is highly efficient once the posterior is obtained. Generating each 1,000-quarter trajectory requires milliseconds on modern hardware. The synthetic data generation completes in minutes, after which we have a reusable training corpus that can be combined with real data under different mixing ratios. This computational efficiency makes the approach practical for iterative experimentation and hyperparameter tuning.

\section{Transformer Likelihood}\label{sec:transformer}

Transformers learn to forecast by treating time series as a kind of language. Just as LLMs predict the next word given preceding text, our model predicts next period's economic conditions given the most recent history. The analogy runs deeper. Words map to discrete tokens, economic variables map to discrete bins; sentence structure captures semantic relationships, temporal patterns capture dynamic relationships. This section explains how we adapt the transformer architecture to macroeconomic forecasting, translating continuous time series into a learnable vocabulary while respecting the multivariate structure of economic data.

From an econometric perspective, the transformer can be interpreted as a nonlinear function estimating dynamic conditional expectations $\mathbb{E}[y_{t+1}\mid y_{1:t}]$ through stacked attention and feedforward layers. Each layer expands the functional space over which expectations are formed, analogous to progressively richer lag polynomials or state transition expansions in classical time series models. The architecture is thus a high-dimensional analogue of recursive forecasting models, but with data-driven lag selection via attention.

\subsection{Tokenization Strategy}

Continuous macroeconomic time series present fundamental challenges to direct application of transformer models, which traditionally operate on discrete tokens from finite vocabularies. Standard approaches like uniform binning or simple discretization fail to capture the complex, time-varying distributions characteristic of economic data. We address this through a percentile-based tokenization scheme that preserves distributional information while enabling efficient sequence modeling.

Beyond enabling transformer architectures, tokenization serves as a principled noise reduction mechanism, which is of critical importance in noise-heavy social science applications. In practice, one does not require the precision implied by continuous measurements. A central bank deciding on interest rates cares whether inflation is broadly low, moderate, or high, not whether it is precisely 2.3\% versus 2.35\%. Our tokenization formalizes this intuition by discretizing continuous observations into economically meaningful bins, effectively filtering out noise while retaining the signal relevant for forecasting and decision-making.

We implement the variable-specific tokenization as follows. For each economic variable $k \in \{1,\ldots,K\}$, we define a partition of the real line
\begin{gather*}
    \mathbb{R}=\bigcup_{j=0}^{J-1} B_{kj}, \qquad B_{kj}\bigcap B_{k\ell}=\emptyset \ \ (j\neq \ell).
\end{gather*}
Here the intervals $\{B_{kj}\}$ are induced by the empirical percentiles $\{p_0, p_{10}, \ldots, p_{90}, p_{100}\}$ computed from the training data (both real and synthetic), which define $J=10$ bins in our baseline specification. Each continuous observation $y_t^{(k)}$ is then mapped to a discrete bin index $z_t^{(k)}$
\begin{gather}
    T_k:\mathbb{R}\to\{0,\ldots,J-1\}, \qquad z_t^{(k)}=T_k\bigl(y_t^{(k)}\bigr),\label{bin}
\end{gather}
where token 0 represents values in the bottom percentile and token $J-1$ represents the top percentile. The multivariate token at time $t$ is then $z_t=(z_t^{(1)},\ldots,z_t^{(K)})$. This percentile-based discretization shares conceptual similarities with approaches in financial time series modeling such as \cite{kronos}, though our multivariate embedding strategy discussed below differs from the univariate setting.

This approach has several advantages over uniform binning. First, it ensures balanced token frequencies, as each bin contains approximately 10\% of observations by construction. Second, it automatically handles different scales and distributions across variables. Inflation rate and output growth, for example, have vastly different ranges. Third, it preserves information about tail events, which are particularly important for macroeconomic forecasting during crisis periods. Lastly, the coarse discretization (ten bins) acts as a regularizer, preventing the model from overfitting to measurement errors or spurious precision in the data, which is particularly valuable given the small sample sizes typical in social science datasets. Our tokenization ensures that tokens carry meaningful economic information about relative magnitudes and tail events while suppressing irrelevant high-frequency variation.

\subsection{Model Architecture}

Our architectural design incorporates two key adaptations for macroeconomic forecasting. First, we train seven specialized transformers rather than one monolithic model, inspired by modular systems in autonomous driving where separate networks handle distinct tasks (e.g., Tesla's Full Self-Driving uses over 100 specialized transformers). Each transformer sees the latest multivariate history but predicts just one target variable in the next period. This design avoids the curse of dimensionality. Jointly predicting seven variables would require modeling the combinatorial space of all possible outcome combinations.

\begin{figure}[!tb]
    \centering
    \vspace{0.2cm}
    \begin{tikzpicture}[scale=0.9, transform shape, >=latex]
    \tikzstyle{data} = [circle, draw=black, thick, fill=white, minimum size=1.1cm, align=center, font=\small]
    \tikzstyle{embed} = [rectangle, draw=darkblue, fill=darkblue!20, minimum width=0.6cm, minimum height=1.5cm]
    \tikzstyle{arrow} = [->, thick, black]

    \node[anchor=west, font=\bfseries\small] at (-8, 5) {1. Raw Data};
    \node[data] (d1) at (-3, 5) {2.3\%\\\scriptsize Output};
    \node[data] (d2) at (0, 5) {1.8\%\\\scriptsize Inflation};
    \node (d3) at (2, 5) {$\cdots$};
    \node[data] (d4) at (4, 5) {4.2\%\\\scriptsize Interest};
    
    \node[anchor=west, font=\bfseries\small] at (-8, 2.5) {2. Tokenization};
    
    \foreach \i in {0,...,9} {
        \draw[fill=white, draw=gray!50, thin] (-4.0 + \i*0.2, 2.4) rectangle (-4.0 + \i*0.2 + 0.2, 2.6);
    }
    \draw[black, thin] (-4.0, 2.4) rectangle (-2.0, 2.6);
    \draw[fill=darkblue!60, draw=black] (-4.0 + 5*0.2, 2.4) rectangle (-4.0 + 5*0.2 + 0.2, 2.6);
    \coordinate (t1_in) at (-3, 2.6);
    \coordinate (t1_out) at (-3, 2.4);

    \foreach \i in {0,...,9} {
        \draw[fill=white, draw=gray!50, thin] (-1.0 + \i*0.2, 2.4) rectangle (-1.0 + \i*0.2 + 0.2, 2.6);
    }
    \draw[black, thin] (-1.0, 2.4) rectangle (1.0, 2.6);
    \draw[fill=darkblue!60, draw=black] (-1.0 + 3*0.2, 2.4) rectangle (-1.0 + 3*0.2 + 0.2, 2.6);
    \coordinate (t2_in) at (0, 2.6);
    \coordinate (t2_out) at (0, 2.4);

    \node at (2, 2.5) {$\cdots$};

    \foreach \i in {0,...,9} {
        \draw[fill=white, draw=gray!50, thin] (3.0 + \i*0.2, 2.4) rectangle (3.0 + \i*0.2 + 0.2, 2.6);
    }
    \draw[black, thin] (3.0, 2.4) rectangle (5.0, 2.6);
    \draw[fill=darkblue!60, draw=black] (3.0 + 7*0.2, 2.4) rectangle (3.0 + 7*0.2 + 0.2, 2.6);
    \coordinate (t4_in) at (4, 2.6);
    \coordinate (t4_out) at (4, 2.4);
    
    \draw[arrow] (d1) -- node[right, font=\scriptsize, xshift=1mm, black] {$y_t^{(1)}$} (t1_in);
    \draw[arrow] (d2) -- node[right, font=\scriptsize, xshift=1mm, black] {$y_t^{(2)}$} (t2_in);
    \draw[arrow] (d4) -- node[right, font=\scriptsize, xshift=1mm, black] {$y_t^{(K)}$} (t4_in);
    
    \node[anchor=west, font=\bfseries\small] at (-8, 0.0) {3. Embedding};
    
    \def\mx{-3}
    \draw[fill=white, draw=black, thin] (\mx-1.0, -0.75) rectangle (\mx+1.0, 0.75);
    \foreach \j in {1,...,9} { \draw[gray!50, thin] (\mx-1.0 + \j*0.2, -0.75) -- (\mx-1.0 + \j*0.2, 0.75); }
    \draw[fill=darkblue!60, draw=black] (\mx-1.0 + 5*0.2, -0.75) rectangle (\mx-1.0 + 6*0.2, 0.75);
    \coordinate (e1_top) at (\mx, 0.75);
    \coordinate (e1_bottom) at (\mx, -0.75);

    \def\mx{0}
    \draw[fill=white, draw=black, thin] (\mx-1.0, -0.75) rectangle (\mx+1.0, 0.75);
    \foreach \j in {1,...,9} { \draw[gray!50, thin] (\mx-1.0 + \j*0.2, -0.75) -- (\mx-1.0 + \j*0.2, 0.75); }
    \draw[fill=darkblue!60, draw=black] (\mx-1.0 + 3*0.2, -0.75) rectangle (\mx-1.0 + 4*0.2, 0.75);
    \coordinate (e2_top) at (\mx, 0.75);
    \coordinate (e2_bottom) at (\mx, -0.75);

    \node at (2, 0.0) {$\cdots$};

    \def\mx{4}
    \draw[fill=white, draw=black, thin] (\mx-1.0, -0.75) rectangle (\mx+1.0, 0.75);
    \foreach \j in {1,...,9} { \draw[gray!50, thin] (\mx-1.0 + \j*0.2, -0.75) -- (\mx-1.0 + \j*0.2, 0.75); }
    \draw[fill=darkblue!60, draw=black] (\mx-1.0 + 7*0.2, -0.75) rectangle (\mx-1.0 + 8*0.2, 0.75);
    \coordinate (e4_top) at (\mx, 0.75);
    \coordinate (e4_bottom) at (\mx, -0.75);
    
    \draw[arrow] (t1_out) -- node[right, font=\scriptsize, xshift=1mm, black] {$z_t^{(1)}$} (e1_top);
    \draw[arrow] (t2_out) -- node[right, font=\scriptsize, xshift=1mm, black] {$z_t^{(2)}$} (e2_top);
    \draw[arrow] (t4_out) -- node[right, font=\scriptsize, xshift=1mm, black] {$z_t^{(K)}$} (e4_top);
    
    \node[anchor=west, font=\bfseries\small] at (-8, -2.5) {4. Concatenation};
    
    \draw[fill=darkblue!60, draw=darkblue] (-3.5, -2.7) rectangle (-1.1, -2.3);
    \coordinate (c1_top) at (-2.3, -2.3);
    
    \draw[fill=darkblue!60, draw=darkblue] (-1.1, -2.7) rectangle (1.3, -2.3);
    \coordinate (c2_top) at (0.1, -2.3);
    
    \node at (1.8, -2.5) {$\cdots$};
    
    \draw[fill=darkblue!60, draw=darkblue] (2.3, -2.7) rectangle (4.7, -2.3);
    \coordinate (c4_top) at (3.5, -2.3);
    
    \draw[arrow] (e1_bottom) -- node[right, font=\scriptsize, xshift=1mm, black] {$h_t^{(1)}$} (c1_top);
    \draw[arrow] (e2_bottom) -- node[right, font=\scriptsize, xshift=1mm, black] {$h_t^{(2)}$} (c2_top);
    \draw[arrow] (e4_bottom) -- node[right, font=\scriptsize, xshift=1mm, black] {$h_t^{(K)}$} (c4_top);
    
\end{tikzpicture}
    \vspace{0.2cm}
    \caption{From continuous data to concatenated embeddings. The process transforms raw economic indicators (top) into a unified vector representation (bottom) suitable for the transformer. Each variable is independently discretized into tokens, mapped to a dense vector via its own embedding table, and finally concatenated to form the state representation.}\label{fig:embedding}
\end{figure}

The second architectural adaptation addresses multivariate embedding. Transformers operate on continuous vector representations rather than discrete tokens directly. The embedding table converts each discrete token into a continuous vector in high-dimensional space, allowing the model to learn relationships and patterns through numerical operations. This is analogous to how economists use VARs to capture multivariate dynamics. However, naively creating one embedding table for all possible token combinations across seven variables would require $10^7 = 10$ million entries, which is difficult to estimate even with ample synthetic data. Instead, we use separate embedding tables for each variable (70 embeddings total). Specifically, if $E_k\in\mathbb{R}^{d\times J}$ denotes the embedding matrix for variable $k$ and $e_j$ is the $j$th canonical basis vector, then each token $z_t^{(k)}$ is mapped to a dense embedding vector
\begin{gather}
    h_t^{(k)} = E_k e_{z_t^{(k)}} \in \mathbb{R}^{d}, \qquad k=1,\ldots,K,\label{embedding}
\end{gather}
where $d=8$ is the embedding dimension in our baseline specification. These variable-specific embeddings are then concatenated to form the multivariate state representation
\begin{gather}
    h_t = \left[h_t^{(1)'}, \ldots, h_t^{(K)'}\right]\in \mathbb{R}^E\label{concatenation}
\end{gather}
with total embedding dimension $E=Kd=56$. Equations \eqref{bin}--\eqref{concatenation} are illustrated schematically in Figure \ref{fig:embedding}. This strategy reduces the number of parameters by several orders of magnitude while respecting that the same token index can represent distinct economic states for different variables.

\begin{table}[!tb]
    \centering
    \caption{Model configuration}
    \label{tab:config}
    \begin{tabular}{lcc}
        \toprule
        Hyperparameter & Notation & Value \\
        \midrule
        Embedding dimension & $E$ & 56 \\
        Number of layers & $L$ & 2 \\
        Number of attention heads & $D$ & 1 \\
        Sequence length (quarters) & $T$ & 4 \\
        Number of variables & $K$ & 7 \\
        Tokens per variable & $J$ & 10 \\
        Batch size & $B$ & 256 \\
        Real data mixing ratio & $\alpha$ & 0.1 \\
        \midrule
        Total parameters & & 52,538 \\
        \bottomrule
    \end{tabular}
\end{table}

Our compact transformer processes sequences of length $T=4$ (analogous to four lags in quarterly VARs) of concatenated embeddings $\{h_{t-T+1}, \ldots, h_t\}$ to predict the next-period token for the target variable. The architecture uses $L=2$ transformer layers with $D=1$ attention head per layer, totaling approximately 52,000 parameters (see Table \ref{tab:config}). The attention mechanism learns to weight past states adaptively. During recessions, the model attends to previous crisis episodes and leading indicators like investment; during expansions, attention focuses on recent dynamics. This resembles how economists perform historical decompositions in DSGE models, but the attention weights are learned from data rather than imposed by theoretical restrictions.

We optimize using cross-entropy loss on next-token prediction, the discrete analogue of maximizing one-step-ahead likelihood. This training objective directly parallels maximum likelihood estimation in standard econometrics. Just as ordered probit or logit models estimate conditional probabilities for discrete outcomes, our transformer learns the probability distribution over the $J=10$ ordered tokens for the target variable. The key difference lies in the functional form. Rather than assuming parametric link functions with linear indices, the transformer uses stacked attention and deep feedforward layers to approximate the conditional distribution nonparametrically. This flexibility allows the model to capture complex, high-dimensional dependencies that would be difficult to specify in traditional discrete choice frameworks.

\subsection{Mixed Training}

How should we combine 231 historical observations with millions of synthetic trajectories when training the transformer? The synthetic data vastly outnumber real observations, yet both contain essential information. Simulations encode theoretical structure while historical data capture actual patterns.

Standard transfer learning strategies prove inadequate in this asymmetric data regime. The conventional approach of pretraining on abundant data, then fine-tuning on limited samples, assumes that the abundant data provide general representations that require only minor adjustments. But with overwhelming synthetic abundance versus scarce real data, fine-tuning would barely shift pretrained representations. The model effectively memorizes DSGE dynamics while ignoring empirical signals, analogous to an overconfident prior that resists updating despite 
empirical evidence.

We instead implement \textit{mixed-batch training}, a protocol that treats theory and data as complements rather than substitutes. Each training iteration constructs mini-batches of size $B=256$ by randomly sampling observations from both sources: 10\% from real data and 90\% from synthetic DSGE simulations in our baseline specification. This ensures real observations continuously influence learning from the first iteration, while synthetic data provide stable gradients, prevent overfitting to the small historical sample, and expose the model to rare events (crises, regime changes, tail realizations) underrepresented in 77 years of the U.S. quarterly data.

The mixing ratio $\alpha \in [0,1]$ serves as an interpretable hyperparameter controlling the theory-data balance. From a Bayesian perspective, $\alpha$ functions analogously to prior precision or effective sample size. It determines the relative weight assigned to structural beliefs encoded in DSGE simulations versus empirical evidence in historical observations. At $\alpha=0$, the model learns pure DSGE dynamics, effectively imposing the structural model as a dogmatic prior. At $\alpha=1$, the transformer sees only real data and risks overfitting, similar to flat-prior estimation with minimal regularization. Our baseline $\alpha=0.1$ strikes a middle ground. The DSGE model provides a theory-informed prior through synthetic trajectories, while historical observations serve as the likelihood that updates beliefs. Formally, letting $P_{\mathrm{data}}$ denote the empirical distribution of real sequences and $P_{\mathrm{sim}}$ the distribution induced by DSGE simulations, we estimate the collection of transformer parameters $\vartheta$ for each economic variable $k$ by minimizing
\begin{gather*}
    \alpha\,\mathbb{E}_{P_{\mathrm{data}}}\!\left[-\log p_\vartheta\bigl(z_{t+1}^{(k)}\mid z_{1:t}\bigr)\right]
+
(1-\alpha)\,\mathbb{E}_{P_{\mathrm{sim}}}\!\left[-\log p_\vartheta\bigl(z_{t+1}^{(k)}\mid z_{1:t}\bigr)\right].
\end{gather*}
This objective makes explicit that training combines a data likelihood term and a simulation-based regularization term in a single criterion.

\section{Posterior Results}\label{sec:posterior}

Computationally, our approach is efficient in that training completes in minutes on standard hardware with parallel processing (MacOS with Apple M2 Silicon chip), enabling rapid experimentation with different transformer architectures and DSGE specifications. This accessibility contrasts sharply with LLM training, which typically requires industrial-scale computational resources. The efficiency stems from our compact architecture (approximately 50,000 parameters) and the highly structured nature of macroeconomic data, where theoretical constraints dramatically reduce the effective dimensionality of the learning problem.

\subsection{Forecasting Performance}

\begin{figure}[!tb]
    \centering
    \includegraphics[width=0.96\textwidth]{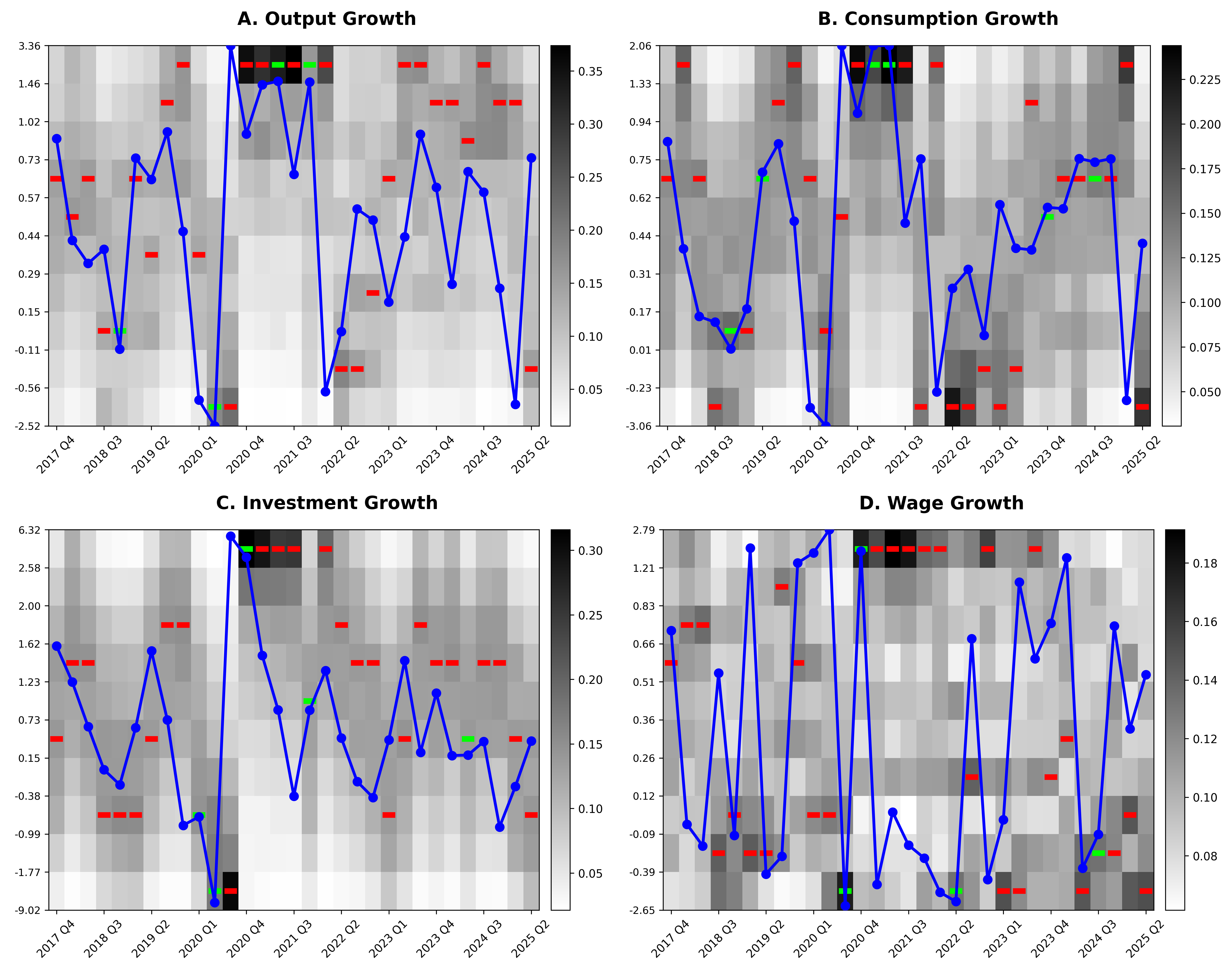}
    \caption{Out-of-sample predictions for the four growth-rate variables, 2017:Q4--2025:Q2. Notes: The grayscale heatmap shows the full predictive token distribution at each time step for each variable. The blue line with circles represents the actual observed values, mapped to their corresponding positions on the token scale. Horizontal bars indicate the model's point predictions. Green bars denote correct predictions, while red bars indicate prediction errors. The vertical axis shows the percentile bin boundaries estimated from the training data.}
    \label{fig:growth}
\end{figure}

\begin{figure}[!tb]
    \centering
    \includegraphics[width=0.96\textwidth]{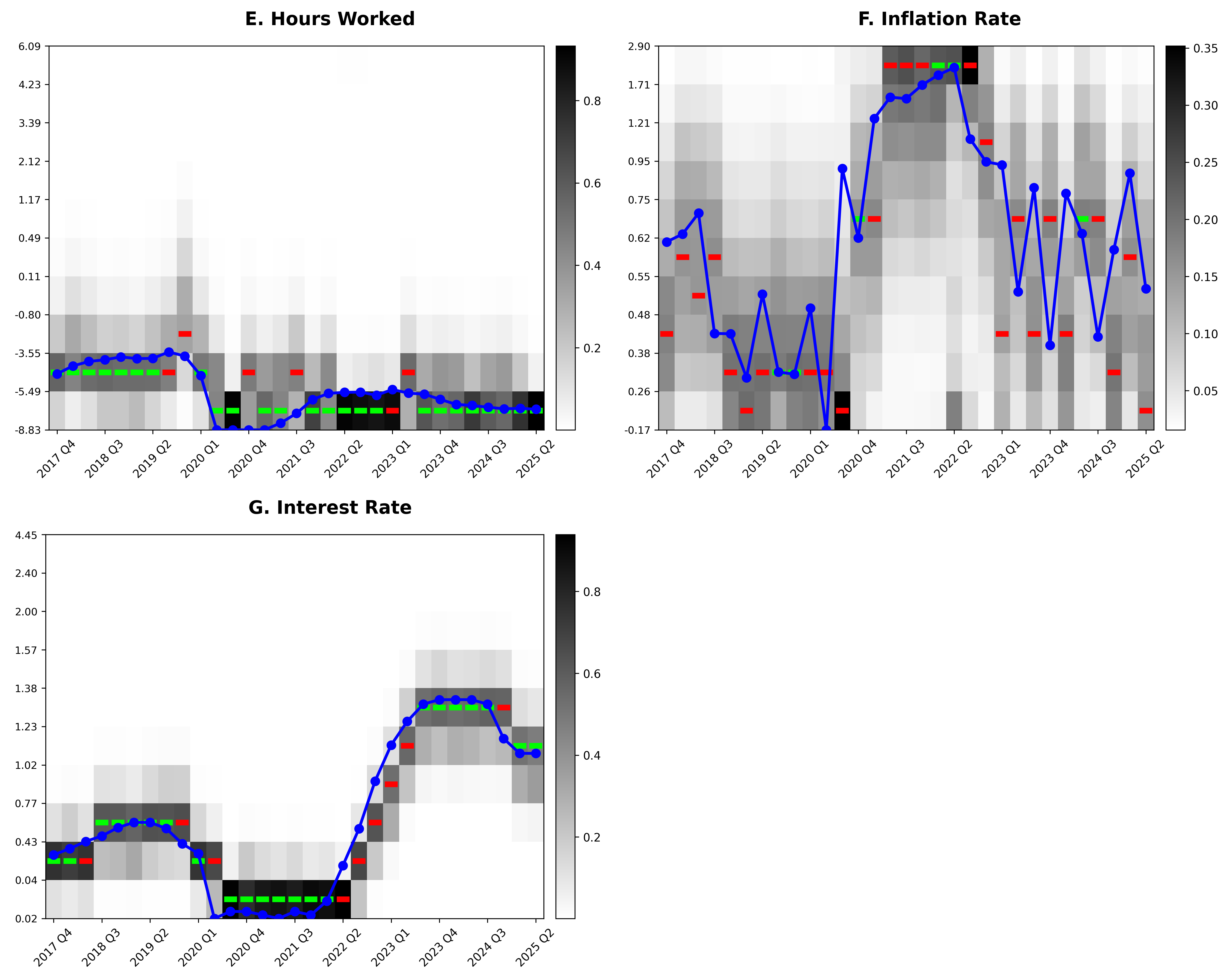}
    \caption{Out-of-sample predictions for the three level variables, 2017:Q4--2025:Q2. Notes: The grayscale heatmap shows the full predictive token distribution at each time step for each variable. The blue line with circles represents the actual observed values, mapped to their corresponding positions on the token scale. Horizontal bars indicate the model's point predictions. Green bars denote correct predictions, while red bars indicate prediction errors. The vertical axis shows the percentile bin boundaries estimated from the training data.}
    \label{fig:level}
\end{figure}

Figures~\ref{fig:growth} and~\ref{fig:level} display three complementary views of the model's forecasting performance for the growth and level variables, respectively. The grayscale heatmap shows the full predictive token distribution at each time step for each variable. This probabilistic output distinguishes our approach from traditional point forecasting methods. The model naturally quantifies its uncertainty by assigning probabilities across the entire support of possible outcomes. During periods of high confidence, the distribution concentrates sharply on a narrow range of tokens (appearing as dark vertical bands), while during uncertain periods, the probability mass spreads more evenly across multiple tokens. The blue line with circles represents the actual observed values, mapped to their corresponding positions on the token scale. Horizontal bars indicate the model's point predictions. Green bars denote correct predictions where the predicted token matches the actual token, while red bars indicate prediction errors. The vertical axis shows the percentile bin boundaries estimated from the training data.

Several patterns emerge from the forecasting results. First, the model performs better on persistent level variables (hours worked, inflation, interest rate) than on volatile growth rate variables (output, consumption, investment). This is consistent with theoretical expectations. Level variables exhibit stronger autocorrelation and mean reversion, providing more predictable dynamics. Growth rates, by contrast, reflect period-to-period changes that are inherently noisier and harder to forecast. The transformer appears to internalize this hierarchy of persistence, similar to the eigenvalue structure embedded in linearized DSGE models.

Second, rather than making large forecasting mistakes that would indicate random guessing or complete model failure, errors typically involve confusing neighboring states, for instance, predicting moderate inflation when actual inflation is slightly high. This ordinal structure suggests the model has learned economically sensible rankings of macroeconomic conditions, even when precise point predictions deviate from realized values. From a policy perspective, this pattern is encouraging since decision-makers care more about correctly identifying the broad state of the economy (such as expansion versus recession or low versus high inflation) than about predicting values with spurious precision.

These results validate our mixed training approach. The synthetic DSGE simulations provide abundant training signal about economic co-movements and dynamic relationships, including crisis scenarios and tail events rarely observed in historical data, while the real observations anchor predictions to actual patterns. The 10\% vs. 90\% mixture proves effective. The model generalizes well to out-of-sample periods, suggesting it has learned structured economic relationships from simulations that transfer to real forecasting tasks.

\subsection{Comparison with VAR}

\begin{table}[!tb]
    \centering
    \caption{VAR(4) versus transformer with strong DSGE prior ($\alpha=0.1$)}
    \label{tab:var_benchmark}
    \begin{threeparttable}
    \small
    \begin{tabular}{lcccc}
        \toprule
        & \multicolumn{2}{c}{Token accuracy} & \multicolumn{2}{c}{Predictive log likelihood} \\
        \cmidrule(ll){2-3} \cmidrule(lr){4-5}
        Variable & VAR(4) & Transformer & VAR(4) & Transformer \\
        \midrule
        Output growth      & $0.065$ & $0.129^*$ & $-3.398\ $ & $-2.283^*$ \\
        Consumption growth & $0.161$ & $0.194^*$ & $-2.933\ $ & $-2.251^*$ \\
        Investment growth  & $0.161$ & $0.161\ $  & $-3.051\ $ & $-2.150^*$ \\
        Wage growth        & $0.129$ & $0.129\ $ & $-2.103^*$ & $-2.325\ $ \\
        Hours worked       & $0.806$ & $0.806\ $ & $-0.461^*$ & $-0.696\ $ \\
        Inflation          & $0.194$ & $0.194\ $ & $-3.707\ $ & $-2.010^*$ \\
        Interest rate      & $0.484$ & $0.710^*$ & $-1.588\ $ & $-0.869^*$ \\
        \bottomrule
    \end{tabular}
    \begin{tablenotes}[flushleft]
        \item Notes: The evaluation sample is 2017:Q4--2025:Q2. Asterisk ($*$) denotes better relative performance. Token accuracy is the share of correct decile-bin predictions. Predictive log likelihood is the average one-step-ahead log probability assigned to the realized bin. The random benchmark is $0.100$ for accuracy and $\log(0.1) = -2.303$ for predictive log likelihood.
    \end{tablenotes}
    \end{threeparttable}
\end{table}

Table~\ref{tab:var_benchmark} reports the out-of-sample comparison between the transformer and a rolling one-step-ahead VAR(4) benchmark under the baseline specification with a strong DSGE prior ($\alpha=0.1$). We report both token accuracy and average predictive log likelihood over the decile bins induced by the transformer tokenization. On token accuracy, the transformer performs at least as well as the VAR benchmark, matching it on four variables and improving upon it on output growth, consumption growth, and interest rate. The largest gain occurs for the interest rate forecast, where accuracy increases from 0.484 under VAR(4) to 0.710 under the transformer. On predictive log likelihood, the transformer outperforms the VAR on five of the seven variables and exceeds the random baseline on six series, whereas the VAR does so on only three. The VAR retains a slight advantage only for wage growth and hours worked. These findings suggest that, under a strong DSGE prior, the transformer provides reliable and in several cases superior distributional forecasts relative to a standard linear benchmark.

\subsection{DSGE Prior Weight}

The baseline results above use a mixed-batch ratio of $\alpha=0.1$, so that real data account for 10\% of each batch and DSGE-generated synthetic data account for the remaining 90\%. To assess the role of the DSGE prior, we consider a more data-heavy specification with $\alpha=0.9$, where real data account for 90\% of the batch and synthetic data only 10\%. This experiment weakens the influence of theory-consistent simulations and places much greater weight on the limited historical sample.

\begin{table}[!tb]
    \centering
    \caption{VAR(4) versus transformer with weak DSGE prior ($\alpha=0.9$)}
    \label{tab:var_benchmark_alpha09}
    \begin{threeparttable}
    \small
    \begin{tabular}{lcccc}
        \toprule
        & \multicolumn{2}{c}{Token accuracy} & \multicolumn{2}{c}{Predictive log likelihood} \\
        \cmidrule(lr){2-3} \cmidrule(lr){4-5}
        Variable & VAR(4) & Transformer & VAR(4) & Transformer \\
        \midrule
        Output growth      & $0.065\ $ & $0.129^*$ & $-3.398\ $ & $-2.328^*$ \\
        Consumption growth & $0.161\ $ & $0.194^*$ & $-2.933\ $ & $-2.301^*$ \\
        Investment growth  & $0.161\ $ & $0.161\ $ & $-3.051\ $ & $-2.184^*$ \\
        Wage growth        & $0.129^*$ & $0.065\ $ & $-2.103^*$ & $-2.333\ $ \\
        Hours worked       & $0.806^*$ & $0.774\ $ & $-0.461^*$ & $-0.559\ $ \\
        Inflation          & $0.194^*$ & $0.161\ $ & $-3.707\ $ & $-2.214^*$ \\
        Interest rate      & $0.484\ $ & $0.581^*$ & $-1.588\ $ & $-1.195^*$ \\
        \bottomrule
    \end{tabular}
    \begin{tablenotes}[flushleft]
        \item Notes: The evaluation sample is 2017:Q4--2025:Q2. Asterisk ($*$) denotes better relative performance. Token accuracy is the share of correct decile-bin predictions. Predictive log likelihood is the average one-step-ahead log probability assigned to the realized bin. The random benchmark is $0.100$ for accuracy and $\log(0.1) = -2.303$ for predictive log likelihood.
    \end{tablenotes}
    \end{threeparttable}
\end{table}

Table~\ref{tab:var_benchmark_alpha09} shows that weakening the DSGE prior leads to a deterioration in the transformer performance relative to the strong-prior baseline. On token accuracy, the transformer continues to outperform VAR(4) for output growth, consumption growth, and interest rate, and matches it for investment growth. However, its performance weakens for wage growth, hours worked, and inflation. Relative to the $\alpha=0.1$ benchmark, the broad pattern is that relaxing DSGE restrictions reduces the transformer's overall forecasting accuracy, even though some individual series remain competitive.

The predictive log likelihood results share a similar pattern. Under $\alpha=0.9$, the transformer still outperforms VAR(4) on five of the seven variables, but its gains are generally smaller than under the strong DSGE prior, and in several cases the predictive distributions move closer to the random benchmark of $-2.303$. This deterioration in performance is due to overfitting: when training relies too heavily on the short historical sample, the model can no longer use DSGE-generated trajectories as an effective source of regularization. One noteworthy exception is hours worked: although the transformer still underperforms VAR(4) on this series, its predictive log likelihood improves from $-0.696$ under the strong prior to $-0.559$ under the weak prior. This suggests that relaxing DSGE restrictions may help by mitigating misspecification in the structural simulator.

Taken together, these findings continue to support the interpretation of mixed-batch training as a theory-data tradeoff. The synthetic DSGE data do not merely enlarge the training sample mechanically; they inject economically coherent structure that stabilizes learning and improves out-of-sample generalization. A strong DSGE prior remains important for the transformer's best overall performance, but the hours-worked exception suggests that the appropriate prior strength may vary across variables and may also reflect limitations of the DSGE model itself.

\subsection{Robustness to Finer Tokenization}

The benchmark results in Table~\ref{tab:var_benchmark} use a decile-based tokenization with $J=10$ bins per variable. To assess whether the comparison with VAR depends on this relatively coarse discretization, we re-estimate the evaluation metrics under a finer partition with $J=20$ bins, while keeping the strong DSGE prior with $\alpha=0.1$. Moving from 10 to 20 bins makes exact classification more demanding for both models, since each realized outcome is assigned to a narrower interval.

\begin{table}[!tb]
    \centering
    \caption{VAR(4) versus transformer with finer tokenization ($J=20$)}
    \label{tab:var_benchmark_20bins}
    \begin{threeparttable}
    \small
    \begin{tabular}{lcccc}
        \toprule
        & \multicolumn{2}{c}{Token accuracy} & \multicolumn{2}{c}{Predictive log likelihood} \\
        \cmidrule(lr){2-3} \cmidrule(lr){4-5}
        Variable & VAR(4) & Transformer & VAR(4) & Transformer \\
        \midrule
        Output growth      & $0.065\ $ & $0.097^*$ & $-4.063\ $ & $-2.990^*$ \\
        Consumption growth & $0.161^*$ & $0.129\ $ & $-3.515\ $ & $-2.969^*$ \\
        Investment growth  & $0.032\ $ & $0.065^*$ & $-3.596\ $ & $-2.916^*$ \\
        Wage growth        & $0.065\ $ & $0.097^*$ & $-2.756^*$ & $-2.985\ $ \\
        Hours worked       & $0.645^*$ & $0.581\ $ & $-0.813^*$ & $-1.229\ $ \\
        Inflation          & $0.065\ $ & $0.065\ $ & $-4.494\ $ & $-2.660^*$ \\
        Interest rate      & $0.419\ $ & $0.548^*$ & $-2.041\ $ & $-1.346^*$ \\
        \bottomrule
    \end{tabular}
    \begin{tablenotes}[flushleft]
        \item Notes: The evaluation sample is 2017:Q4--2025:Q2. Asterisk ($*$) denotes better relative performance. Token accuracy is the share of correct decile-bin predictions. Predictive log likelihood is the average one-step-ahead log probability assigned to the realized bin. The random benchmark is $0.050$ for accuracy and $\log(0.05) = -2.996$ for predictive log likelihood.
    \end{tablenotes}
    \end{threeparttable}
\end{table}

Table~\ref{tab:var_benchmark_20bins} shows that the main message is robust to a finer tokenization scheme. On token accuracy, the transformer's advantage becomes less uniform, which is expected because exact-hit prediction is more stringent with 20 bins than with 10. Even so, the transformer continues to outperform VAR(4) for output growth, investment growth, wage growth, and interest rate, ties on inflation, and underperforms only for consumption growth and hours worked.

The predictive log likelihood comparison remains more favorable to the transformer. It outperforms VAR(4) on five of the seven variables, with gains for output growth, consumption growth, investment growth, inflation, and interest rate. These results suggest that, even when exact token prediction becomes harder under a finer discretization, the transformer continues to assign more probability mass around the realized outcomes for most series. The main exceptions are wage growth and hours worked, where the VAR retains an advantage in predictive log likelihood.

Overall, the 20-bin results indicate that the superiority of the transformer in distributional forecasting is not an artifact of the baseline decile partition. A finer discretization narrows the margin on exact classification, but the transformer remains competitive on token accuracy and retains a broad advantage in predictive likelihood under the strong DSGE prior.

\subsection{Robustness to Network Depth}

The baseline specification in Table~\ref{tab:var_benchmark} uses a compact transformer with $L=2$ layers and roughly 50,000 parameters. To assess whether the comparison with VAR depends on this relatively compact architecture, we re-estimate the benchmark using a deeper transformer with $L=4$ layers, which increases the parameter count to just over 100,000 while leaving the tokenization scheme ($J=10$) and strong DSGE prior ($\alpha=0.1$) unchanged.

\begin{table}[!tb]
    \centering
    \caption{VAR(4) versus deeper transformer with four layers}
    \label{tab:var_benchmark_4layers}
    \begin{threeparttable}
    \small
    \begin{tabular}{lcccc}
        \toprule
        & \multicolumn{2}{c}{Token accuracy} & \multicolumn{2}{c}{Predictive log likelihood} \\
        \cmidrule(lr){2-3} \cmidrule(lr){4-5}
        Variable & VAR(4) & Transformer & VAR(4) & Transformer \\
        \midrule
        Output growth      & $0.065\ $ & $0.161^*$ & $-3.398\ $ & $-2.296^*$ \\
        Consumption growth & $0.161\ $ & $0.226^*$ & $-2.933\ $ & $-2.212^*$ \\
        Investment growth  & $0.161^*$ & $0.129\ $ & $-3.051\ $ & $-2.167^*$ \\
        Wage growth        & $0.129\ $ & $0.129\ $ & $-2.103^*$ & $-2.253\ $ \\
        Hours worked       & $0.806^*$ & $0.710\ $ & $-0.461^*$ & $-0.720\ $ \\
        Inflation          & $0.194\ $ & $0.290^*$ & $-3.707\ $ & $-1.966^*$ \\
        Interest rate      & $0.484\ $ & $0.677^*$ & $-1.588\ $ & $-0.904^*$ \\
        \bottomrule
    \end{tabular}
    \begin{tablenotes}[flushleft]
        \item Notes: The evaluation sample is 2017:Q4--2025:Q2. Asterisk ($*$) denotes better relative performance. Token accuracy is the share of correct decile-bin predictions. Predictive log likelihood is the average one-step-ahead log probability assigned to the realized bin. The random benchmark is $0.100$ for accuracy and $\log(0.1) = -2.303$ for predictive log likelihood.
    \end{tablenotes}
    \end{threeparttable}
\end{table}

Table~\ref{tab:var_benchmark_4layers} shows that the main message is robust to a substantially deeper network. On token accuracy, the deeper transformer outperforms VAR(4) for output growth, consumption growth, inflation, and interest rate, ties on wage growth, and underperforms only for investment growth and hours worked. The largest gains again appear for persistent nominal variables such as inflation and interest rate.

The predictive log likelihood results remain broadly favorable to the transformer. The deeper specification outperforms VAR(4) on five of the seven variables, with improvements for output growth, consumption growth, investment growth, inflation, and interest rate. As in the baseline specification, the main exceptions are wage growth and hours worked, where the VAR retains an advantage.

Overall, increasing the network depth from two to four layers, thereby doubling the parameter count, does not alter the qualitative comparison with VAR. The transformer continues to deliver a broad advantage in distributional forecasting while remaining competitive on token accuracy.

\subsection{Discussions}

Our findings carry several implications for economic forecasting practice. First, transformer architectures, appropriately adapted, can deliver strong predictive performance despite small samples. The key innovation is treating DSGE models not as competing forecasting methods but as complementary data augmentation mechanisms that encode economic theory.

Second, the strong performance of the baseline compact architecture (two network layers, 56-dimensional embeddings, approximately 50,000 parameters), together with similar qualitative results from a deeper four-layer specification with more than 100,000 parameters, suggests that macroeconomic forecasting may not require the massive models common in transformer applications. Economic data are highly structured with well-understood theoretical constraints. This structure allows relatively small and moderate-sized models to learn effectively, particularly when training incorporates theory-consistent simulations. The computational accessibility facilitates reproducibility and broader adoption in academic and policy settings.

Third, the framework admits natural extensions. Multi-horizon forecasting could be implemented by training separate models for different forecast horizons or by extending the architecture to predict multiple steps ahead. Scenario analysis under alternative policy rules could leverage the DSGE simulator to generate counterfactual training data. Ensemble approaches combining multiple DSGE specifications could provide robustness to model misspecification. These extensions could yield macro-aligned forecasting systems that serve as economic counterparts to language-foundation architectures.

Finally, the approach enables novel applications beyond pure forecasting. The framework can be interpreted as an amortized likelihood learner. Once trained, the transformer approximates $p(y_{t+1}\mid y_{1:t},\theta)$ for draws from the DSGE parameter posterior, effectively learning a fast surrogate for the computationally expensive DSGE likelihood. Future work could embed this approach within a full Bayesian workflow, using the transformer as a differentiable likelihood approximation for simulation-based inference and thereby unifying DSGE estimation and deep forecasting.

\section{Conclusion}\label{sec:conclusion}

We show that a transformer can learn important features of the macroeconomic language when using theory-consistent simulation in the training process. By tokenizing economic variables and training on mixed batches of DSGE-generated synthetic data and real observations, we develop a forecaster that combines structural discipline with predictive flexibility. Relative to a rolling VAR(4) benchmark, the resulting model matches or exceeds token accuracy for all variables and delivers higher predictive likelihood for most series. These findings suggest that theory-infused transformers can usefully complement standard linear benchmarks by delivering more reliable distributional forecasts for macroeconomic variables. More broadly, the same strategy applies to economic settings where structural models are available but data are scarce.

Our approach also speaks to a broader limitation emerging in contemporary large language model development. Scaling laws appear to be reaching diminishing returns despite massive increases in parameters and training data. In contrast, humans learn efficiently by integrating new information with prior structure in a manner resembling Bayesian updating. This suggests that progress may depend less on brute-force scaling and more on domain-specific adaptations. Just as autonomous vehicles and robotic systems learn from physics-based simulators, time series transformers can learn from DSGE-generated synthetic trajectories, opening the door to analogous applications of theory generated synthetic data in other disciplines.

\newpage

\renewcommand{\appendixpagename}{\large\sc\color{darkblue} Appendix}\pdfbookmark[1]{Appendix}{sec:app}
\makeatletter
\def\toclevel@subsection{2}
\makeatother
\numberwithin{equation}{subsection}

\begin{appendices}

\section{Transformer Model}
\label{app:transformer}

This appendix provides a complete and explicit mathematical description of the transformer model
used in the paper. 

\subsection{Observed Data and Conditioning Object}

Let
\[
\{x_t\}_{t=1}^T, \qquad x_t=(x_{t1},\ldots,x_{tK})\in\mathbb{R}^K,
\]
denote the observed multivariate time series. Fix a context length $L\ge 1$. The statistical object
modeled is the conditional distribution
\[
p_\vartheta(x_{t+1}\mid x_t,x_{t-1},\ldots,x_{t-L+1}),
\]
where $\vartheta$ collects all unknown parameters of the model. The conditional distribution is approximated after discretization of the state space, described next.

\subsection{Tokenization (Deterministic Preprocessing)}

For each variable $k\in\{1,\ldots,K\}$, fix an integer $J_k\ge 2$ and a measurable partition
\[
\mathbb{R}=\bigcup_{j=1}^{J_k} B_{kj}, \qquad B_{kj}\bigcap B_{k\ell}=\emptyset \;\; (j\neq \ell).
\]
Define the tokenization map
\[
T_k:\mathbb{R}\to\{1,\ldots,J_k\}, \qquad
T_k(x)=j \;\;\text{if and only if}\;\; x\in B_{kj}.
\]
Let
\[
z_{tk}=T_k(x_{tk}), \qquad z_t=(z_{t1},\ldots,z_{tK}).
\]
The token sequence $\{z_t\}$ is fully observed and deterministic given $\{x_t\}$.

\subsection{Embedding Matrices and Embedded States}

For each variable $k$, define an embedding dimension $d\ge 1$ and an embedding matrix
\[
E_k \in \mathbb{R}^{d\times J_k}.
\]
Each column of $E_k$ is an unknown parameter vector to be estimated. Let $e_j$ denote the $j$th
canonical basis vector in $\mathbb{R}^{J_k}$. The embedded representation of variable $k$ at time $t$
is
\[
h_{tk}^{(0)} = E_k e_{z_{tk}} \in \mathbb{R}^d.
\]
The embedded state vector at time $t$ is
\[
h_t^{(0)} = \sum_{k=1}^K h_{tk}^{(0)} \in \mathbb{R}^d.
\]
For a context window of length $L$, define the stacked embedded state
\[
H_t^{(0)} =
\begin{pmatrix}
h_{t-L+1}^{(0)} \\
\vdots \\
h_t^{(0)}
\end{pmatrix}
\in \mathbb{R}^{L\times d}.
\]

\subsection{Transformer Layers}

Fix an integer $M\ge 1$ denoting the number of transformer layers. For each layer
$\ell\in\{1,\ldots,M\}$, define parameter matrices
\[
W_Q^{(\ell)}, W_K^{(\ell)}, W_V^{(\ell)}, W_O^{(\ell)} \in \mathbb{R}^{d\times d}.
\]
Given the input $H^{(\ell-1)}\in\mathbb{R}^{L\times d}$, define
\[
Q^{(\ell)} = H^{(\ell-1)} W_Q^{(\ell)}, \quad
K^{(\ell)} = H^{(\ell-1)} W_K^{(\ell)}, \quad
V^{(\ell)} = H^{(\ell-1)} W_V^{(\ell)}.
\]
The self-attention operator is
\[
A^{(\ell)} =
\mathrm{softmax}\!\left(\frac{Q^{(\ell)} K^{(\ell)\top}}{\sqrt d}\right) V^{(\ell)},
\]
where $\mathrm{softmax}$ is applied row-wise. The output of layer $\ell$ is
\[
H^{(\ell)} =
\phi\!\left(A^{(\ell)} W_O^{(\ell)} + H^{(\ell-1)}\right),
\]
where $\phi(\cdot)$ is a fixed elementwise nonlinearity (e.g.\ ReLU or GELU). Set
$H^{(0)}=H_t^{(0)}$. After $M$ layers, define
\[
h_t = \text{last row of } H^{(M)} \in \mathbb{R}^d.
\]

\subsection{Output Layer and Conditional Probabilities}

For each variable $k$, define an output weight matrix
\[
U_k \in \mathbb{R}^{J_k\times d}.
\]
The vector of logits for variable $k$ at time $t+1$ is
\[
\ell_{t+1,k} = U_k h_t \in \mathbb{R}^{J_k}.
\]
The conditional probability mass function is
\[
p_\vartheta(z_{t+1,k}=j \mid z_{t:t-L+1})
=
\frac{\exp(\ell_{t+1,k,j})}
{\sum_{m=1}^{J_k}\exp(\ell_{t+1,k,m})}.
\]
The joint conditional distribution factorizes as
\[
p_\vartheta(z_{t+1}\mid z_{t:t-L+1})
=
\prod_{k=1}^K p_\vartheta(z_{t+1,k}\mid z_{t:t-L+1}).
\]

\subsection{Likelihood and Estimation}

Define the log-likelihood contribution at time $t$ as
\[
\ell_t(\vartheta)
=
\sum_{k=1}^K \log p_\vartheta(z_{t+1,k}\mid z_{t:t-L+1}).
\]
The full parameter vector is
\[
\vartheta =
\left(
\{E_k\}_{k=1}^K,
\{W_Q^{(\ell)},W_K^{(\ell)},W_V^{(\ell)},W_O^{(\ell)}\}_{\ell=1}^M,
\{U_k\}_{k=1}^K
\right).
\]
The estimator is defined as the maximum likelihood estimator
\[
\hat\vartheta
=
\arg\max_{\vartheta}
\sum_{t=L}^{T-1} \ell_t(\vartheta).
\]

\subsection{Mixed Training with Simulated Data}

Let $P_{\text{data}}$ denote the empirical distribution of token sequences and
$P_{\text{sim}}$ the distribution induced by DSGE simulations. Estimation minimizes
\[
\alpha\,\mathbb{E}_{P_{\text{data}}}\!\left[-\ell_t(\vartheta)\right]
+
(1-\alpha)\,\mathbb{E}_{P_{\text{sim}}}\!\left[-\ell_t(\vartheta)\right],
\qquad \alpha\in(0,1),
\]
with respect to $\vartheta$. All parameters listed above are estimated jointly by this
criterion.

\section{Python Code}\label{app:code}

We provide open-source Python implementation of our forecasting framework, adapting the codebase from \cite{nanoGPT}, to facilitate replication and extension by researchers. The codebase comprises three modular components designed for clarity and flexibility: \texttt{model.py}, \texttt{train.py}, and \texttt{config.py}. This architecture separates model specification, training logic, and hyperparameter configuration, enabling rapid experimentation without modifying the core functionality.
\begin{itemize}
    \item \textbf{Model architecture (\texttt{model.py})}. This module implements the transformer architecture using PyTorch, a popular deep learning framework widely adopted in machine learning research. The implementation includes: (i) variable-specific embedding tables that map discrete tokens to continuous vectors; (ii) multi-head self-attention layers with causal masking to prevent information leakage from future time steps; (iii) feedforward networks with activation functions and layer normalization; and (iv) an output head producing probability distributions over tokens via softmax activation. The modular design allows researchers to experiment with architectural variations by modifying isolated components rather than rewriting the entire model.
    \item \textbf{Training pipeline (\texttt{train.py})}. This module orchestrates the mixed-batch training protocol. Core functionality includes: (i) data loading routines that read real and synthetic datasets, apply standardization transformations, and implement the percentile-based tokenization scheme; (ii) mixed-batch sampling that constructs training batches by drawing $\alpha \cdot B$ observations from real data and $(1-\alpha) \cdot B$ from synthetic simulations; (iii) optimization loop using Adam optimizer with cross-entropy loss, computing gradients and updating model parameters; and (iv) evaluation routines that compute forecast accuracy metrics on held-out test data and save trained model checkpoints. The implementation handles edge cases such as small real data samples through careful batch construction and supports early stopping based on validation performance.
    \item \textbf{Configuration (\texttt{config.py})}. This module centralizes all hyperparameters and data paths in a single location. Key settings include: (i) architectural parameters ($E$, $L$, $H$, $T$) controlling model capacity; (ii) training parameters (batch size $B$, mixing ratio $\alpha$, learning rate, number of epochs); (iii) tokenization settings (number of bins $J$, percentile computation method); and (iv) file paths for real and synthetic datasets. Researchers can modify these settings without touching the model or training code, which facilitates systematic hyperparameter searches and sensitivity analysis.
\end{itemize}

Practitioners can apply the framework to their forecasting problems by preparing two datasets: (i) real data in CSV format with columns for each variable and rows for time periods, and (ii) synthetic data from estimated structural models in the same format. The training script automatically handles data preprocessing, standardization, and tokenization. After training, the saved model can generate predictive token distributions for each target variable. The complete implementation, including detailed documentation, usage examples, and the exact configurations used in this paper, is available upon request. 

\end{appendices}

\section*{Data Availability}

The data and code used in this paper are available at \href{https://github.com/econdojo/dsgeGPT}{https://github.com/econdojo/dsgeGPT}.

\section*{Generative AI Disclosure}

In preparing the manuscript, the authors used GPT-5.4 for language editing and formatting assistance. The tool was only used to help revise prose for clarity and grammar. All substantive research decisions, model design, estimation, empirical analysis, and final verification of the manuscript and code were carried out by the authors.

\section*{Disclosure Statement}

The authors report no relevant financial or non-financial competing interests.

\bibliographystyle{econometrica}
\bibliography{OneBibToRuleThemAll}

\end{document}